\documentclass[final,5p,times,twocolumn]{elsarticle}
\usepackage{graphicx}
\usepackage{amssymb,amsmath}

\journal{Journal of Physics and Chemistry of Solids}
\begin{document}

\begin{frontmatter}

\title{Importance of Itinerancy and Quantum Fluctuations for the Magnetism in Iron Pnictides}

\author[1,2]{Yu-Zhong Zhang\corref{cor1}}\ead{yzhang@itp.uni-frankfurt.de}
\author[1]{Hunpyo Lee}
\author[1]{Ingo Opahle}
\author[1]{Harald O. Jeschke}
\author[1]{Roser Valent\'\i}

\address[1]{Institut f\"ur Theoretische Physik, Goethe-Universit\"at Frankfurt, Max-von-Laue-Stra{\ss}e 1, 60438 Frankfurt am Main, Germany}
\address[2]{Department of Physics, Tongji University, Shanghai, 200092 P. R. China}

\cortext[cor1]{Corresponding author.}

\begin{abstract}
By applying density functional theory, we find strong evidence for an itinerant nature of magnetism in two families of iron
pnictides. Furthermore, by employing
dynamical mean field theory with continuous time quantum Monte Carlo
as an impurity solver, we observe that the antiferromagnetic metal
with small magnetic moment naturally arises out of coupling between
unfrustrated and frustrated bands. Our results point to a possible
scenario for magnetism in iron pnictides where magnetism originates
from a strong instability at the momentum vector ($\pi$, $\pi$, $\pi$) while
it is reduced by quantum fluctuations due to the coupling between
weakly and strongly frustrated bands.
\end{abstract}


\end{frontmatter}

\section{Introduction}
\label{section:one}

Since the discovery of high-T$_c$ superconductivity in
LaOFeAs~\cite{Kamihara2008},
great effort has been devoted
to pursuing higher superconducting transition temperatures
$T_c$~\cite{Ren2008} as well as to the understanding of the various
phase transitions in these materials as a function of temperature,
pressure and
doping~\cite{Luetkens2009,Kimber2009,Medvedev2009}. Density functional
theory (DFT) calculations~\cite{Boeri2008,Boeri2010} show that
electron-phonon coupling -- though non-negligible -- is not strong
enough to explain the high $T_c$ observed in these systems. Instead,
magnetically mediated pairing has been proposed to account for the
superconducting state due to its proximity to a stripe-type
antiferromagnetic (AF) metallic state~\cite{Singh2008,Mazin2008}.

However, the origin of the stripe-type AF metal, i.e., whether it
comes from the itinerant nature of the Fermi surface (FS)
nesting~\cite{Mazin2008,Cvetkovic09,YZZhang2009,Yaresko2009,YZZhang2010} or a
localized picture of exchange interactions between local
spins~\cite{Si2008,Yildirim2008,Schmidt09}, is still under debate. The
scenario of FS nesting was severely challenged by recent DFT
calculations~\cite{Moon2009}, which found a disconnection between Fe
moment and FS nesting. However, a recent revision of the nature of
magnetism in the iron pnictides -- also within DFT, but with more
precise optimized structures -- has reestablished the close
connection between itinerancy and magnetism~\cite{YZZhang2010}. In
this work, we will present further clear evidence of the itinerant
nature of magnetism by performing DFT calculations for a few
families of iron pnictides.

On the other hand, the disagreement~\cite{Opahle2009,Mazin2008PRB}
of the magnitude of the Fe moment in the stripe-type AF metal
between experimental observations and DFT calculations based on
experimental structures is still controversely discussed. Various
mechanisms have been proposed from DFT calculations, for example,
negative effective on-site electronic interaction
$U$~\cite{Nakamura2008,Ferber2010}, a low moment solution within
GGA+U stabilized by the formation of magnetic
multipoles~\cite{Cricchio2009,comment} or frustration between local
spins~\cite{Si2008,Yildirim2008,Ma2008,Han2009}. Recently, also
mean-field calculations based on a five-band Hubbard model with
positive $U$ obtained a small magnetic moment comparable to
experimental results~\cite{Kaneshita2009,Bascones2010,Brydon2010}. However, the existence of various sets of DFT-derived hopping parameters~\cite{Miyake2010,Graser2009,Kuroki2008} casts doubt on the reliability of the model parameters used in the model calculations.

From DFT calculations it is well known~\cite{Singh2008,Opahle2009,Mazin2008PRB}
that the physical properties of the iron pnictides are highly sensitive to
details of the structure as well as to details of the exchange and
correlation (XC) functionals. For instance,
structural optimization within the local spin density approximation (LSDA)
leads to almost perfect agreement~\cite{Opahle2009} with the most recent
experimental value of the Fe moment~\cite{Qureshi10}.
This indicates, that quantum fluctuations, which are only insufficiently incorporated in the
common approximations to DFT, could strongly improve the agreement with
experiment.

To further explore this, we employ dynamical mean-field theory
(DMFT)~\cite{Georges1996} combined with continuous time quantum
Monte Carlo (CTQMC)~\cite{Rubtsov2005} simulations. A feature that
is common to different sets of DFT-derived hopping parameters is the
presence of some strongly and some weakly frustrated Fe~$3d$
bands~\cite{Miyake2010,Graser2009,Kuroki2008}. In the present work
we consider a {\it
  minimal} two-band Hubbard model which should capture this feature by
considering one band with frustration and the second one without
frustration and we investigate the effect of the coupling of these
two bands on the magnetism of the system. We would like to remark
that while a realistic description of the Fe systems needs at least
a five-band model, the present two-band model should be sufficient
for the proposed analysis. We find that an AF metallic state is
present in a wide range of the interaction parameter $U$ when one
band is highly frustrated and the second one unfrustrated, while the
state is absent when both bands are equally
frustrated~\cite{Lee2009}.

Our results from DFT and DMFT suggest that a strong instability at a momentum vector ($\pi$, $\pi$, $\pi$) is the
promising mechanism for the itinerant magnetism observed in iron
pnictides while quantum fluctuations originating from  the coupling
between weakly and strongly frustrated bands reduce the magnetic
moment and make it more comparable to the experimental observations.

\section{Method and Model}
\label{section:two}

In order to quantify the FS nesting and hence the itinerant nature of the magnetism, we calculate
 (i) the $\mathbf{q}$-dependent Pauli susceptibility at $\omega$=0 with the
constant matrix element approximation, defined as
\begin{equation}
\chi _0\left( \mathbf{q}\right) =-\sum\limits_{\mathbf{k}\alpha \beta }\frac{%
f\left( \varepsilon _{\mathbf{k}\alpha }\right) -f\left( \varepsilon _{%
\mathbf{k+q}\beta }\right) }{\varepsilon _{\mathbf{k}\alpha }-\varepsilon _{%
\mathbf{k+q}\beta }+i\delta }  \label{Paulisusceptibility}
\end{equation}
where $\alpha$ and $\beta$ are band indexes and $q$ and $k$ are
momentum vectors in the Brillouin zone, and (ii) the $k_z$ dispersion
of the FS. These calculations were performed with the full potential
linearized augmented plane wave method as implemented in the WIEN2k
code~\cite{WIEN} with $RK_\text{max}=7$. While 40000~$\mathbf{k}$
points are used in calculating the $k_z$ dispersion of the FS, a three
dimensional grid of $128\times128\times128$ $\mathbf{k}$ and
$\mathbf{q}$ points are employed for the susceptibility. All
calculations were performed in the scalar relativistic approximation.

In order to perform the DFT analysis, (i) we use the available
experimental structures for LaOFeAs~\cite{Nomura2008},
BaFe$_2$As$_2$~\cite{Huang08} and (ii) we obtain fully optimized
structures for BaFe$_{2-x}$Co$_x$As$_2$ within the virtual crystal
approximation and for a few hypothetical compounds like LaOFeSb and
BaFe$_2$Sb$_2$ using the Car-Parrinello~\cite{CarParrinello}
projector-augmented wave~\cite{Bloechl} method. Our optimized
structures compare well with the experimental
ones~\cite{YZZhang2010,Thirupathaiah2010}.  Part of our results are
double-checked by the full potential local orbital (FPLO)
method~\cite{FPLO}. Results are consistent among these methods.
Throughout the paper, the Perdew-Burke-Ernzerhof generalized
gradient approximation (GGA) to DFT has been used.

For our model calculations we consider the following
 two-band Hubbard model
\begin{equation}\begin{split}
  H &=-\sum_{\langle ij\rangle m\sigma}  t_m c^{\dagger}_{im\sigma}
  c_{jm\sigma} -\sum_{\langle ij'\rangle m\sigma}  t_m^{\prime} c^{\dagger}_{im\sigma}
  c_{j'm\sigma} \\
    &+\,U\sum_{im}n_{im\uparrow} n_{im\downarrow} +\sum_{i\sigma\sigma'}\big(U'-\delta_{\sigma \sigma'} J_z\big)n_{i1\sigma}
  n_{i2\sigma'}\,,
\label{eq:hamiltonian}
\end{split}\end{equation}
where $t_m$ ($t_m^{\prime }$) is the intra-band hopping integral
between nearest-neighbor (next nearest-neighbor) sites with band
indices $m=1,2$. For simplification, we neglect inter-band
hybridizations. $U $, $U^{\prime }$ and $J_z$ are the intra-band,
inter-band Coulomb interaction and Ising-type Hund's coupling,
respectively. In our calculations we set $U^{\prime }=\frac U2$ and
$J_z=\frac U4$ which fulfills the rotational invariance condition of
$U=U^{\prime }+2J$ and ignore the spin-flip and pair-hopping
processes. The operators are written in the standard notation of the
multi-band Hubbard model~\cite{Lee2009}. In order to solve this
model we employ the two-sublattice DMFT method~\cite{Georges1996}
which includes the local quantum fluctuation effects and can account
for the AF state, combined with CTQMC
simulations~\cite{Rubtsov2005}. Existing DMFT studies combining with
DFT calculations are focused on paramagnetic
state~\cite{Haule2008,Craco2008,Skornyakov2009,Ishida2010,Aichhorn2009}.
Our calculations are performed on the Bethe lattice.

\section{Results and Discussion}
\label{section:three}

\begin{figure}[htbp]
\includegraphics[width=0.48\textwidth]{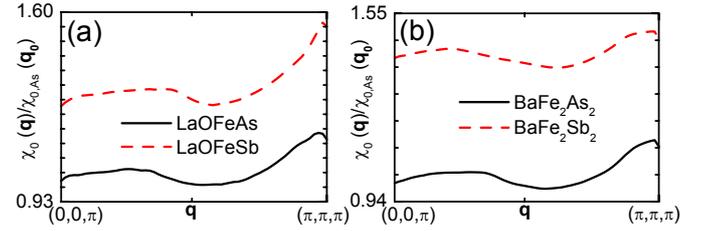}
\caption{(Color online) Comparison of normalized
  $\mathbf{q}$-dependent Pauli susceptibilities calculated within GGA
  at fixed $q_z=\pi$ along the $[110]$ direction between arsenide and
  antimonide of (a) 1111 compounds and (b) 122 compounds. The
  normalization factors are the susceptibilities of the corresponding
  arsenide systems for each type of compound at
  $\mathbf{q}_0=(0,0,\pi)$. Please note that the peak position is not
  exactly at $\mathbf{q}_{\pi}=(\pi,\pi,\pi)$ since the electron and
  hole FSs are nearly nested rather than perfectly nested.}
\label{fig:SusqzPiqxqy}
\end{figure}

\begin{table}[!ht]
\caption{Magnetic moment calculated within spin-polarized GGA based
on
  experimental structures of LaOFeAs and BaFe$_2$As$_2$ and optimized
  structures of the hypothetical compounds LaOFeSb and
  BaFe$_2$Sb$_2$.} \label{table1}
\begin{tabular}{c|cc|cc}
\hline\hline & LaOFeAs & LaOFeSb & BaFe$_2$As$_2$ & BaFe$_2$Sb$_2$
\\ \hline $m(\mu_B)$ & 1.8 & 2.2 & 2.0 & 2.5 \\ \hline
\end{tabular}
\end{table}

In Fig.~\ref{fig:SusqzPiqxqy}, we present the comparison of
$\mathbf{q}$-dependent Pauli susceptibilities between arsenide and
antimonide 1111 and 122 compounds normalized by the susceptibilities
of the corresponding arsenide systems for each type of compound at
$\mathbf{q} _0=(0,0,\pi)$. Here we only consider $q_z=\pi$ since in
both LaOFeAs and BaFe$_2$As$_2$ the Fe spins are AF ordered along
$c$ as experimentally observed and we show the results for
$q_x=q_y$. We observe that in all these four compounds a strong peak
around $\mathbf{q}_\pi=(\pi,\pi,\pi)$ is found, which supports the
presence of stripe-type magnetically ordered states as observed
experimentally. Most importantly, in contrast to an earlier DFT
study~\cite{Moon2009} where the magnetic moment increases as As is
replaced by Sb in both LaOFeAs and BaFe$_2$As$_2$ while the
susceptibilities at $\mathbf{q}_\pi$ decrease in BaFe$_2$Sb$_2$
compared to BaFe$_2$As$_2$, which is interpreted to be evidence of
disconnection between FS nesting and magnetism, our results show
that in both compounds, the susceptibilities at $\mathbf{q}_\pi$
increase as the replacement takes place and simultaneously the
magnetic moment is also enhanced as displayed in
Table.~\ref{table1}. This discrepancy is attributed to the improved
precision of optimized lattice structures in our DFT
calculation~\cite{YZZhang2010}. Therefore, the close connection
between itinerancy and magnetism remains.

\begin{figure}[htbp]
\includegraphics[width=0.48\textwidth]{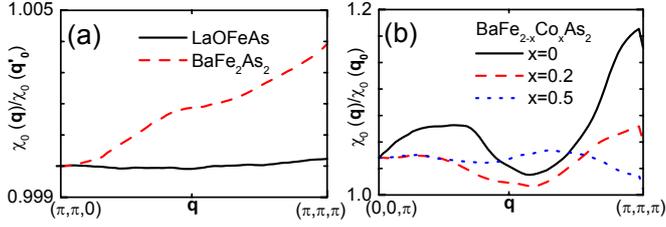}
\caption{(Color online) (a) Normalized $\mathbf{q}$-dependent Pauli
susceptibilities of LaOFeAs and BaFe$_2$As$_2$ at fixed
$q_x=q_y=\pi$ along the $[001]$ direction calculated within GGA. The
normalization factors are the susceptibilities of each compound at
$\mathbf{q}'_0=(\pi,\pi,0)$, and the experimental structure is used.
(b) Normalized $\mathbf{q}$-dependent Pauli susceptibilities of
BaFe$_{2-x}$Co$_x$As$_2$ at fixed $q_z=\pi$ along the $[110]$
direction calculated within GGA. Here $x=0,0.2,0.5$. The
normalization factors are the susceptibilities of each compound at
$\mathbf{q}_0=(0,0,\pi)$.} \label{fig:SusqxPiqyPiqz}
\end{figure}

\begin{figure}[htbp]
\includegraphics[angle=-90,width=0.48\textwidth]{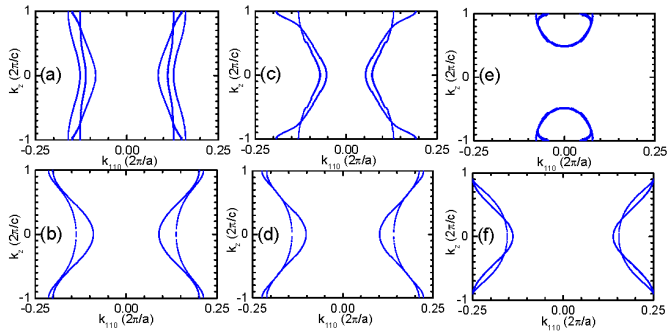}
\caption{(Color online) $k_z$ dispersion of FS calculated within GGA
as a function of Co-doping in BaFe$_{2-x}$Co$_x$As$_2$ around
$\Gamma$ at $x=0$ (a), at $x=0.2$ (c), at $x=0.5$ (e) and around $X$
at $x=0$ (b), at $x=0.2$ (d), at $x=0.5$ (f).}
\label{fig:FSVSdoping}
\end{figure}

In Fig.~\ref{fig:SusqxPiqyPiqz} (a), we show the
$\mathbf{q}$-dependent Pauli susceptibilities of LaOFeAs and
BaFe$_2$As$_2$ at fixed $q_x=\pi$ and $q_y=\pi$, normalized by the
susceptibilities of each compound at $\mathbf{q}'_0=(\pi,\pi,0)$. We
find that while a tiny increase of the susceptibilities is found
from $\mathbf{q}'_0$ to $\mathbf{q}_\pi$ in LaOFeAs, indicating a
dispersionless FS along $c$ and therefore nearly perfect two
dimensional physical properties, a stronger enhancement is seen in
BaFe$_2$As$_2$ suggesting a three dimensional FS topology as shown
in Fig.~\ref{fig:FSVSdoping} (a), (b), even though it is a layered
compound. Such a three dimensionality of the FS has been proposed to
be the mechanism for a nearly isotropic critical field in
(Ba,K)Fe$_2$As$_2$~\cite{Yuan2009}. The common feature in the
susceptibilities for LaOFeAs and BaFe$_2$As$_2$ is that the values
at $\mathbf{q}_\pi$ are larger than those at $\mathbf{q}'_0$ which
can account for the AF arrangement of Fe spins along $c$, although
the interlayer interaction is believed to be small. Such a
consistency again implies a close relation between itinerancy and
magnetism.

It has already been shown that changes in the FS topology can
explain the different nature of structural and magnetic phase
transitions in the 122 compounds under
pressure~\cite{YZZhang2009,YZZhang2010JPCM}. In the following we
will further demonstrate that the phase transitions in
BaFe$_2$As$_2$ under Co-doping are also related to the change of FS
topology. It is known from experiments~\cite{Chu2009} that for
BaFe$_{2-x}$Co$_x$As$_2$ at $x=0$, the system shows a stripe-type AF
metal.  At $x=0.2$ the system becomes superconducting, and finally
at $x=0.5$, both magnetization and superconductivity disappear. In
Fig.~\ref{fig:FSVSdoping} we show the $k_z$ dispersion of the FS for
these three cases around $\Gamma$ (left panels) and $X$ (right
panels). We find that the FS along $c$ becomes more dispersive with
Co doping. At $x=0$ a strong instability in the Pauli susceptibility
-- though the FS nesting is not perfect -- is present at
$\mathbf{q}_\pi$ as shown in Fig.~\ref{fig:SusqxPiqyPiqz} (b) solid
curve, supporting the AF state. At $x=0.2$, the distortion of the FS
indicates a strong suppression of the $\mathbf{q}_\pi$ instability
(see Fig.~\ref{fig:SusqxPiqyPiqz} (b) dashed curve) and therefore of
the magnetization. However, the superconducting state which,
according to the spin fluctuation theory~\cite{MoriyaUeda}, is
related to the instabilities around $\mathbf{q}_\pi$ may in such a
situation be more favourable than magnetization.  At $x=0.5$, the FS
around $\Gamma$ shrinks to a Fermi pocket and is even more distorted
around $X$, suggesting that no obvious instability in the
susceptibility will be present (see Fig.~\ref{fig:SusqxPiqyPiqz} (b)
dotted curve) and leading to the disappearance of both magnetic
ordering and superconductivity.

\begin{figure}[htbp]
\includegraphics[width=0.48\textwidth]{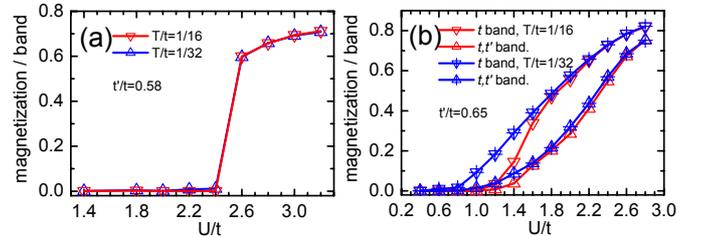}
\caption{(Color online) Magnetization per band of a two-band Hubbard
  model at two temperatures calculated by DMFT(CTQMC) as a function of
  interaction strength $U$ with (a) two bands equally highly
  frustrated and (b) one band highly frustrated, the other
  unfrustrated.}
\label{fig:FMUF}
\end{figure}

After having presented various evidence for the itinerant nature of
magnetism in iron pnictides, we will investigate in what follows a
possible mechanism for the reduced magnetic moment observed
experimentally compared to DFT
calculations~\cite{YZZhang2009,Opahle2009,Mazin2008PRB}. As
mentioned in Section \ref{section:one}, we would like to extract
essential physics from a simplified model as introduced in Section
\ref{section:two}. In this model, local quantum fluctuations are
included in the calculations by employing DMFT. Fig.~\ref{fig:FMUF}
shows the magnetization as a function of $U$ at two temperatures.
Combining these results with the analysis of density of states in
Ref.~\cite{Lee2009}, we conclude that when two bands are equally
highly frustrated (see Fig.~\ref{fig:FMUF} (a)), a first-order phase
transition from a paramagnetic metal to an AF insulator state is
observed and an AF metallic state is absent, while, if one band is
unfrustrated and the second one highly frustrated (see
Fig.~\ref{fig:FMUF} (b)), several continuous phase transitions
appear separately in these two bands, and an AF metal with small
magnetic moment appears. This indicates that an AF metallic state
with small magnetic moment emerges out of the coupling between
highly frustrated and unfrustrated bands, which is the case in iron
pnictides as mentioned in Section \ref{section:one}, rather than out
of a pure frustration effect.

\section{Conclusion}
\label{section:four}

In summary, we presented various evidence for the close relation
between itinerancy and the magnetism observed experimentally. Our
results support the itinerant nature of magnetism in iron pnictides
and suggest that a strong instability at the nesting vector
$\mathbf{q}_\pi$ is responsible for the magnetism in iron pnictides.
We also propose that the reason why the reduced magnetic moment in
the iron pnictides cannot be reproduced by LSDA or GGA calculations
employing the experimental lattice structure is the insufficient
incorporation of quantum fluctuations in such calculations. By
applying the DMFT approach to a simplified two-orbital Hubbard
model, we propose that coupling of strongly frustrated and
unfrustrated bands may be the mechanism for the reduction of the
magnetic moment, rather than a pure frustration effect.

We gratefully acknowledge useful discussions and comments from L.
Nordstr{\"o}m and the Deutsche Forschungsgemeinschaft (DFG) for
financial support through SFB/TRR 49, SPP 1458, Emmy Noether
programs and the Helmholtz Association for support through
HA216/EMMI.


\begin{thebibliography}{99}

\bibitem{Kamihara2008} Y. Kamihara, T. Watanabe, M. Hirano, and
  H. Hosono, J.  Am. Chem. Soc. {\bf 130}, 3296 (2008).

\bibitem{Ren2008} Z. A. Ren, W. Lu, J. Yang, W. Yi, X. L. Shen,
  Z. C. Li, G. C. Che, X. L. Dong, L. L. Sun, F. Zhou, and Z. X. Zhao,
  Chin. Phys. Lett. {\bf 25}, 2215 (2008).

\bibitem{Luetkens2009} H. Luetkens, H.-H. Klauss, M. Kraken, F. J. Litterst, T. Dellmann, R. Klingeler, C. Hess, R. Khasanov, A. Amato, C. Baines, M. Kosmala, O. J. Schumann, M. Braden, J. Hamann-Borrero, N. Leps, A. Kondrat, G. Behr, J. Werner, B. B{\"u}chner, Nature Materials {\bf 8},
305 (2009).

\bibitem{Kimber2009} S. A. J. Kimber, A. Kreyssig, Y.-Z. Zhang, H. O.
  Jeschke, R. Valent{\'\i}, F. Yokaichiya, E. Colombier, J. Yan,
  T. C. Hansen, T. Chatterji, R. J.  McQueeney, P. C. Canfield,
  A. I. Goldman and D. N. Argyriou, Nature Materials {\bf 8}, 471
  (2009).

\bibitem{Medvedev2009} S. Medvedev, T.M. McQueen, I. Trojan, T. Palasyuk, M.I. Eremets, R.J. Cava, S. Naghavi, F. Casper, V. Ksenofontov, G. Wortmann, C. Felser, Nature Materials {\bf 8},
630 (2009).

\bibitem{Boeri2008} L. Boeri, O. V. Dolgov, A. A. Golubov, Phys. Rev. Lett. {\bf 101}, 026403 (2008).

\bibitem{Boeri2010} L. Boeri, M. Calandra, I. I. Mazin, O. V. Dolgov, F. Mauri, arXiv:1004.1943.

\bibitem{Singh2008} D. J. Singh and M.-H. Du, Phys. Rev. Lett. {\bf
  100}, 237003 (2008).

\bibitem{Mazin2008} I. I. Mazin, D. J. Singh, M. D. Johannes and
  M. H. Du, Phys. Rev. Lett. {\bf 101}, 057003 (2008).

\bibitem{Cvetkovic09} V. Cvetkovic and Z. Tesanovic,
  Europhys. Lett. {\bf 85}, 37002 (2009).

\bibitem{YZZhang2009} Y.-Z. Zhang, H. C. Kandpal, I. Opahle,
  H. O. Jeschke, and R.  Valent{\'\i}, Phys. Rev. B {\bf 80}, 094530
  (2009).

\bibitem{Yaresko2009} A. N. Yaresko, G.-Q. Liu, V. N. Antonov, O.K. Andersen, Phys. Rev. B {\bf 79}, 144421
  (2009).

\bibitem{YZZhang2010} Y.-Z. Zhang, I. Opahle,
  H. O. Jeschke, and R.  Valent{\'\i}, Phys. Rev. B {\bf 81}, 094505
  (2010).

\bibitem{Si2008} Q. Si and E. Abrahams, Phys. Rev. Lett. {\bf 101},
  076401 (2008).

\bibitem{Yildirim2008} T. Yildirim, Phys. Rev. Lett. {\bf 101}, 057010
  (2008).

\bibitem{Schmidt09} B. Schmidt, M. Siahatgar, P. Thalmeier,
  Phys. Rev. B {\bf 81}, 165101 (2010).

\bibitem{Moon2009} C.-Y. Moon, S. Y. Park, and H. J. Choi, Phys. Rev. B
  {\bf 80}, 054522 (2009).

\bibitem{Opahle2009} I. Opahle, H. C. Kandpal, Y. Zhang, C. Gros, and R.
  Valent{\'\i}, Phys. Rev. B {\bf 79}, 024509 (2009).

\bibitem{Mazin2008PRB} I. I. Mazin, M. D. Johannes, L. Boeri,
  K. Koepernik, and D. J.  Singh, Phys. Rev. B {\bf 78}, 085104
  (2008).

\bibitem{Nakamura2008} H. Nakamura, N. Hayashi, N. Nakai, M. Machida, arXiv:0806.4804.

\bibitem{Ferber2010} J. Ferber, Y. Z. Zhang, H. O. Jeschke,
  R. Valent{\'\i}, arXiv:1005.1374.

\bibitem{Cricchio2009} F. Cricchio, O. Gr{\aa}n{\"a}s,
  L. Nordstr{\"o}m, Phys. Rev. B {\bf 81}, 140403(R) (2010).

\bibitem{comment} The stabilized solution with small magnetic moment was found in GGA+U with a positive U and employing the
double counting scheme of around mean field. This solution ceases to
be favorable when the double counting scheme of fully localized
limit is considered.

\bibitem{Ma2008} F. Ma, Z.-Y. Lu, T. Xiang, Phys. Rev. B {\bf 78},
  224517 (2008).

\bibitem{Han2009} M. J. Han, Q. Yin, W. E. Pickett, and S. Y. Savrasov,
  Phys. Rev.  Lett. {\bf 102}, 107003 (2009).

\bibitem{Kaneshita2009} E. Kaneshita, T. Morinari, T. Tohyama, Phys. Rev. Lett. {\bf 103},
  247202 (2009).

\bibitem{Bascones2010} E. Bascones, M. J. Calder{\'\o}n, and B. Valenzuela, Phys. Rev. Lett. {\bf 104}, 227201 (2010).

\bibitem{Brydon2010} P. M. R. Brydon, Maria Daghofer, and Carsten Timm, arXiv:1007.1949v1.

\bibitem{Miyake2010} T. Miyake, K. Nakamura, R. Arita, M. Imada, J. Phys. Soc. Jpn. {\bf 79}, 044705 (2010).

\bibitem{Graser2009} S. Graser, T. A. Maier, P. J. Hirschfeld, D. J. Scalapino, New J. Phys. {\bf 11}, 025016 (2009).

\bibitem{Kuroki2008} K. Kuroki, S. Onari, R. Arita, H. Usui, Y. Tanaka, H. Kontani, and H. Aoki, Phys. Rev. Lett. {\bf 101}, 087004 (2008).

\bibitem{Qureshi10} N. Qureshi, Y. Drees, J. Werner, S. Wurmehl, C. Hess,
R. Klingeler, B. B\"uchner, M. T. Fern\'{a}ndez-D\'{i}az, and M. Braden,
arXiv:1002.4326.

\bibitem{Georges1996} A. Georges, G. Kotliar, W. Krauth, M. J. Rozenberg, Rev. Mod. Phys. {\bf 68}, 13 (1996).

\bibitem{Rubtsov2005} A. N. Rubtsov, V. V. Savkin, A. I. Lichtenstein, Phys. Rev. B {\bf 72}, 035122 (2005).

\bibitem{Lee2009} H. Lee, Y.-Z. Zhang, H. O. Jeschke, and
  R. Valent{\'\i}, Phys. Rev. B {\bf 81}, 220506(R) (2010).

\bibitem{WIEN} P. Blaha, K. Schwarz, G. Madsen, D. Kvaniscka, and J. Luitz, WIEN2K, An Augmented Plane
  Wave+Local Orbitals Program for Calculating Crystal, edited by
  K. Schwarz (Techn. University,Vienna, Austria, 2001).

\bibitem{Nomura2008} T. Nomura, S. W. Kim, Y. Kamihara, M. Hirano,
  P. V. Sushko, K. Kato, M. Takata, A. L. Shluger, H. Hosono,
  Supercond. Sci. Technol. {\bf 21}, 125028 (2008).

\bibitem{Huang08} Q. Huang, Y. Qiu, Wei Bao, M. A. Green, J. W. Lynn,
  Y. C.  Gasparovic, T. Wu, G. Wu, and X. H. Chen,
  Phys. Rev. Lett. {\bf 101}, 257003 (2008).

\bibitem{CarParrinello} R. Car, M. Parrinello, Phys. Rev. Lett. {\bf
  55}, 2471 (1985).

\bibitem{Bloechl} P. E. Bl{\"o}chl, Phys. Rev. B {\bf 50}, 17953
  (1994).

\bibitem{Thirupathaiah2010} S. Thirupathaiah, S. de Jong, R. Ovsyannikov, H. A. D{\"u}rr, A. Varykhalov, R. Follath, Y. Huang, R. Huisman, M. S. Golden, Yu-Zhong Zhang, H. O. Jeschke, R. Valent{\'\i}, A. Erb, A. Gloskovskii, and J. Fink, Phys. Rev. B {\bf 81},
104512 (2009).

\bibitem{FPLO} K. Koepernik and H. Eschrig, Phys. Rev. B {\bf 59}, 1743
  (1999).  http://www.FPLO.de

\bibitem{Haule2008} K. Haule, J. H. Shim, G. Kotliar, Phys. Rev. Lett. {\bf 100}, 226402 (2008).

\bibitem{Craco2008} L. Craco, M. S. Laad, S. Leoni, H. Rosner, Phys. Rev. B {\bf 78}, 134511 (2008)

\bibitem{Skornyakov2009} S. L. Skornyakov, A. V. Efremov, N. A. Skorikov, M. A. Korotin, Yu. A. Izyumov, V. I. Anisimov, A. V. Kozhevnikov, D. Vollhardt, Phys. Rev. B {\bf 80}, 092501 (2009)

\bibitem{Ishida2010} H. Ishida, A. Liebsch, Phys. Rev. B {\bf 81}, 054513 (2010)

\bibitem{Aichhorn2009} M. Aichhorn, L. Pourovskii, V. Vildosola, M. Ferrero, O. Parcollet, T. Miyake, A. Georges, S. Biermann, Phys. Rev. B {\bf 80}, 085101 (2009).

\bibitem{Yuan2009} H. Q. Yuan, J. Singleton, F. F. Balakirev, S. A. Baily, G. F. Chen, J. L. Luo, N. L. Wang, Nature {\bf 457}, 565 (2009).

\bibitem{YZZhang2010JPCM} Y.-Z. Zhang, I. Opahle,
  H. O. Jeschke, and R.  Valent{\'\i}, J. Phys.: Condensed Matter {\bf 22}, 164208 (2010).

\bibitem{Chu2009} J.-H. Chu, J. G. Analytis, C. Kucharczyk, and I. R. Fisher, Phys. Rev. B {\bf 79}, 014506 (2009).

\bibitem{MoriyaUeda} T. Moriya, K. Ueda, Rep. Prog. Phys. {\bf 66}, 1299 (2003).

\end{thebibliography}
\end{document}